\documentstyle[pre,aps,epsfig,eqsecnum]{revtex}

\begin{document}
\draft  

\title{Simulated ecology-driven sympatric speciation}
\author{J.S. S\'a Martins$^{\dag}$, S. Moss de Oliveira$^{\ast}$, and 
G.A. de Medeiros$^{\ast}$}
\address{$\dag$ Colorado Center for Chaos and Complexity, CIRES, CB 216,  
University of Colorado, Boulder, Colorado, USA 80309}
\address{$\ast$ Instituto de F\'{\i}sica, Universidade Federal Fluminense, 
Av. Litor\^anea s/n, Boa Viagem, Niter\'oi, 24210-340, RJ, Brazil}

\maketitle

\begin{abstract}
We introduce a multi-locus genetically acquired phenotype, submitted to 
mutations and with selective value, in an age-structured model for biological 
aging. This phenotype describes a single-trait effect of the environment on an 
individual, and we study the resulting distribution of this trait among the 
population. In particular, our simulations show that the appearance of a 
double phenotypic attractor in the ecology induces the emergence of a stable 
polymorphism, as observed in the Galapagos finches. In the presence of this 
polymorphism, the simulations generate short-term speciation, when mating 
preferences are also allowed to suffer mutations and acquire selective value.
\pacs{PACS numbers: 87.23.Cc, 07.05.Tp}
\end{abstract}

\section{Introduction}

The theory of evolution studies processes that are extremely complex and whose 
characteristic times usually extend over many human life spans. Because of 
this inherent complexity, these studies do not lend themselves readily to the 
establishment of discrete, falsifiable hypotheses. For this reason, observations 
and experimental results emanating from evolutionary studies often elicit 
broadly differing interpretations \cite{tauber}. A large number of mathematical 
models have been applied to a variety of evolutionary problems, as well as to 
the related fields of population dynamics and evolutionary ecology, composing 
a unified approach under the name of evolutionary population ecology 
\cite{roughgarden}. The success of many of these models, together with the 
above-mentioned difficulties in the interpretation of observational data, has 
triggered the development of an integrated computational effort for the study 
of these problems. Computer simulations of the behavior of virtual populations, 
immersed in well-defined habitats and evolving under a representation of the 
dynamics of evolution, have generated a wealth of reliable data with which to 
test our models and the assumptions they rely on.

Our approach in this paper is to use a computational representation to study the 
genetical patterns generated by the order-disorder conflict between selective 
pressure and mutation accumulation in the presence of an environment that favors 
particular phenotype configurations. Our model is a variation on the popular Penna 
model \cite{penna}, a bit-string representation of genetic dynamics for age-structured 
populations that has been successfully used to address a variety of problems in 
the field. It provides a simple metaphor for evolutionary dynamics 
in terms of the mutation accumulation theory. In essence, each individual is 
represented by a double strand of inherited genetic material, subject to the 
addition of harmful mutations at birth. Its life span is determined by the 
position of the active deleterious mutations in the genome, while non-mutated 
genes have no effect on the individual adaptation to its environment.

To this underlying age-structured genome we add a non-structured portion, which 
can be seen as representing genes that do not have an age-dependent action. The 
configuration of this new portion of the genome is translated into a phenotype 
which determines, through its match with an environment-dependent ideal, the 
probability of survival when the number of harmful mutations is not critical. 
Models with similar characteristics have been studied recently, but under a 
different perspective \cite{pekalski}. In our case, the phenotype is translated 
into a fitness function merely by counting the number $n$ of positions in the 
bit-string that are set to a given value, say $1$, and the individual fitness is 
a function of this number $f(n)$. This strategy simulates a single phenotypical 
trait which is encoded in a number of genes. We then study the distribution of 
phenotypes of the equilibrium population, identifying its main patterns.

Beyond the theoretical interest of unveiling the features generated by 
evolutionary dynamics on the phenotype of the population, our model also allows us 
to address a much more challenging problem, namely that of sympatric speciation. 
Speciation involves the division of a species on an adaptive peak, so that each 
part moves onto a new adaptive peak without either one going against the 
upward force of natural selection. This process is readily envisioned if a 
species becomes subdivided by a physical barrier, whereby each part experiences 
different mutations, population fluctuations and selective forces, in what is 
called the allopatric model of speciation. In contrast, conceiving the division 
of a single population and radiation onto separate peaks without geographical 
isolation, in what is called sympatric speciation, is intuitively more 
difficult \cite{tauber}. Through which mechanism can a single population of 
interbreeding organisms be converted into two reproductively isolated segments 
in the absence of spatial barriers or hindrances to gene exchange?

As an answer to this question, a general mechanism for sympatric speciation has 
been modeled with a simple one-locus two-allele dynamics that shows how a stable 
polymorphism can exist in a heterogeneous environment with two niches, even when 
the adults form a single randomly-mating population \cite{maynard}. Increasingly 
realistic models that explicitly incorporate density-dependent competition for 
resources among phenotypes have since appeared in the literature \cite{rough}. The 
present paper is another link in this chain. We show that a stable polymorphism 
is generated when the environment switches from a one to a two-niche configuration 
in a model with age-structured genomes and competition among phenotypes, as a 
necessary result of the dynamics of evolution.

The establishment of a stable polymorphism is the first step towards sympatric 
speciation. The second is the subsequent evolution of reproductive isolation 
between the populations in the two niches. The key question here is whether a 
single gene difference could produce selective coefficients large enough to drive 
reproductive isolation. It is generically thought that this isolation requires a 
genetic association between traits conferring adaptation to a niche and traits 
causing mating preferences \cite{grant1}. Alternatively, one can consider 
assortative mating of phenotypes; since a gene for assortative mating can increase 
to fixation sympatrically, speciation could be the outcome of sexual selection 
alone. This is what our model shows: once we introduce a gene which describes the 
mating preferences of an individual, it is lead by the dynamics to a fixation 
pattern in which alike phenotypes mate exclusively.

As a metaphor to the results obtained in our simulations, we compare them 
qualitatively with the observed dynamics of the Galapagos finches. Growing 
evidence for the onset of stable polymorphism in nature comes from years of 
observations of seasonal morphological variations in the population of ground 
finches in the Galapagos Islands. The assumptions that selection, mediated through 
rainfall and its effects on food supply, can have a dramatic impact on finch 
phenotypes, and that much of the morphological variation in these birds is 
genetically inherited are reasonably well established since the field work of P.R. 
Grant and his collaborators \cite{boag,grant,lack}. 

\section{The Penna model with phenotype selection}

In this section we will briefly describe the main characteristics of
the standard age-structured Penna model, together with the features we
added to it to represent phenotypic selection. For a recent review on the
applications of the model we direct the reader to Ref. \cite{school}
and references therein.

For sexual populations, the genome of each organism is represented by
two computer words. In each word, a bit set to 
one at a locus corresponds to a mutated deleterious allele; a ``perfect'' strand 
would be composed solely of zeros. The effect of this mutation may be felt by the 
individual at all ages equal to or above the numerical order of that locus in the 
word. As an example, a bit set to $1$ at the second position of one of the 
bit-strings means that a harmful effect may become present in the life history of 
the organism to which it corresponds after it has lived for two time periods. The 
diploid character of the genome provides for a coverage against the effectiveness 
of these harmful mutations. Thus, a mutation in a position of one of the strands 
is felt as harmful only if this is a homozygote locus or if the mutated allele is 
dominant. In a homozygote locus, a harmful mutation must be present in both strings 
at the same position to be effective. The concept of dominance, on the other hand, 
relates to loci in the genome in which a harmful mutation in just one strand is 
enough to make it affect the organism's life. The life span of an individual is 
controlled by the number of effective mutations active at any instant in time. 
This number must be smaller than a specified threshold to keep the individual 
alive; it dies as soon as this limit is reached. 

An individual may also die because of intra-specific competition for
the finite 
resources of the environment or because of the action of predators. In the 
standard model, these constraints are taken care of by the so-called Verhulst 
factor. This is a logistic-type term that introduces a mean-field random 
death probability, independent of the quality of the genome, and in the 
simulations its main purpose is to limit the size of the population. Its usual 
expression is $V(P) = P/C$, where $P$ is the total population at 
some time step and $C$ is a parameter of the simulation, traditionally 
called the carrying capacity of the environment. This name, inherited
from simple logistic models \cite{roughgarden}, is somewhat misleading
in the context of more sophisticated models; in those simple models
the equilibrium population ends up to be $C$, thus justifying its
name. This is not the case for models that include a more detailed 
microscopic description of the evolutionary interactions, where the 
equilibrium population is a complex function of many different parameters.

Since really random deaths in nature can hardly play any significant
role in population dynamics, this concept has already been criticized
in the literature, and a couple of alternate implementations analyzed 
\cite{cebrat}. Rather than random, one should expect that deaths
caused either by predation or because of intra-specific competition
should also have selective value. The probability of 
their occurrence should be dependent on the fitness of the individual to the 
environment, and this fitness a function of the match between the individual's 
phenotypic expression of genetically acquired traits and a phenotype ideally 
adapted to its habitat. The problems that such a microscopic description of 
selective dynamics entails are extremely complex and are far from
being solved. 
The present paper reports on an attempt to address 
this issue in the context of an age-structured model. In addition to the usual 
double strand of genes with age-dependent expression, we introduce an extra pair 
for each individual in the population. In this computer representation, each genome 
may be considered as composed of $64$ loci - in a $32$-bit word machine - of which 
half have age-dependent expression. The dynamics of reproduction and mutations, to 
be described in what follows, is the same for both the age-structured and the new 
strings. Meiosis, phenotypic expression controlled by homozygose or dominance, 
and the introduction of random mutations at birth affect equally all loci. The new 
portion of the phenotype, however, represents some individual trait, such as the 
size of the beak in a ground finch, for example, that may have selective value. We 
chose a multi loci representation of a single trait for simplicity, and claim that 
there is no loss of generality in this choice for our present purposes. The 
selective value of this phenotype is expressed by a fitness function $f(n)$, 
chosen to depend on a single variable $n$: the number of $1$'s expressed in the 
non-structured portion of the phenotype, either because of homozygose 
($11$ alleles) or the presence of $10$ ($01$) configuration in a locus where the 
$1$ allele is dominant. This fitness function encapsulates the selective value of a 
particular phenotype and restricts the multi-dimensional space of the interaction 
between the individual and its environment to one single dimension. In this 
discretized mathematical model, the probability of death by intra-specific 
competition at each time step is now given by $V(P)/f(n)$. The 
interaction with other individuals is still mediated by the mean-field $V(P)$, 
but its effect is no longer uniform. It depends on the particular genetically 
acquired configuration of each individual and, although stochastic, escapes from 
the biologically unmotivated randomness mentioned above.

\begin{figure}
\centerline{\psfig{file=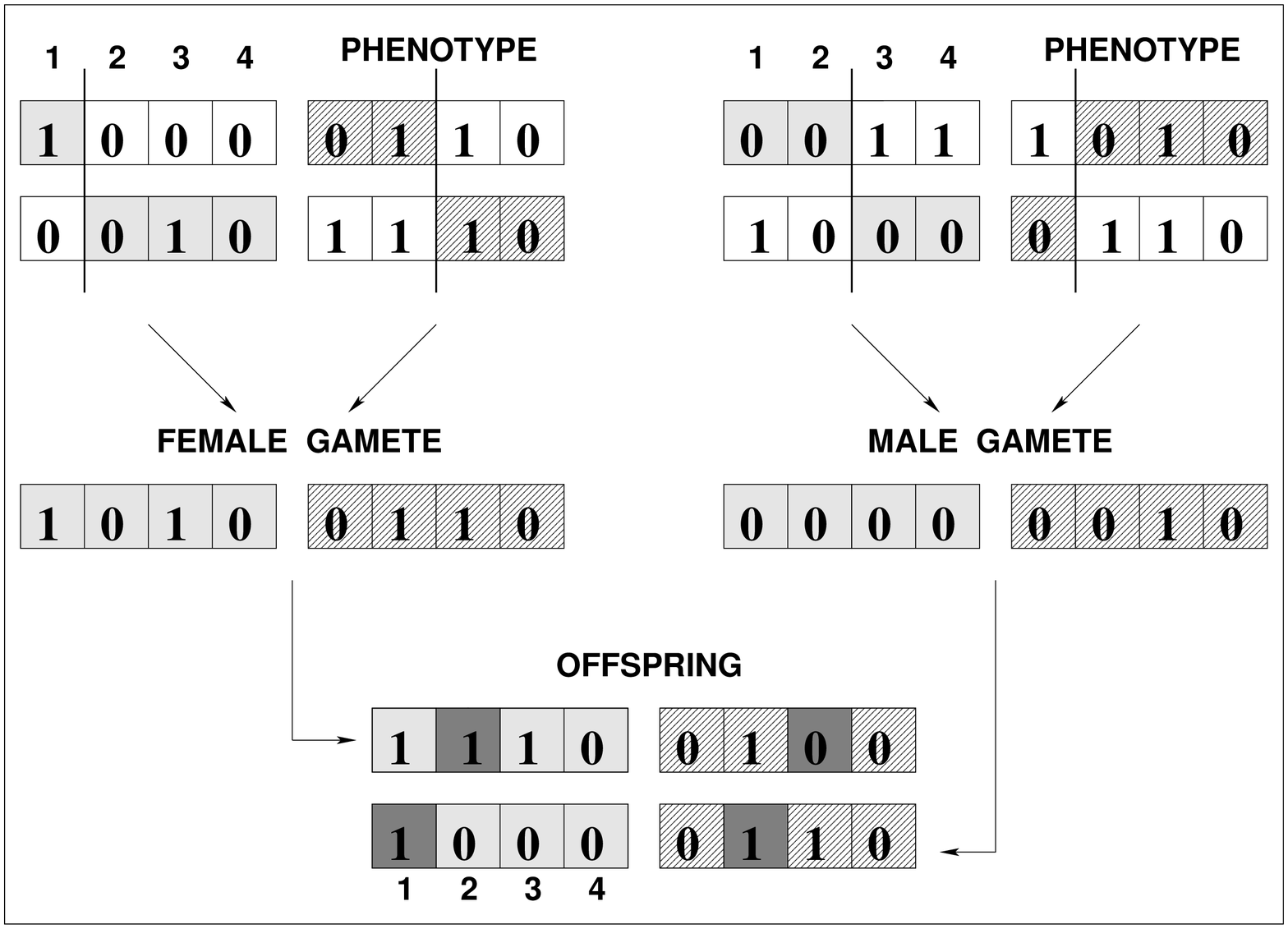,width=14cm,angle=0}}
\caption{Schematical representation of the reproductive process. The age-structured 
part of the gamete has a light-shaded background, while the non-structured part is 
shown over diagonal stripes. The dark-shaded squares of the offspring genome 
correspond to the positions where new mutations were added.}
\label{schema}
\end{figure}

Reproduction is modeled by the introduction of new genomes in the population. Each 
female becomes reproductive after having reached a minimum age, after which she 
generates a fixed number of offspring at the completion of each period of life. 
The meiotic cycle is represented by the generation of a
single-stranded cell out of the diploid genome, now composed of two 
functionally different portions. To do so, each string of the parent
genome is cut at a randomly selected position, the same for both
strings and independently for each functional part, and the left part
of one is combined with the right part of the other, thus generating
two new combinations of the original genes - see Figure
\ref{schema}. The selection of one of these completes the formation of
the haploid gamete coming from the mother. For mating, a male is
randomly selected in the population and undergoes the same meiotic cycle, 
generating a second haploid gamete out of his genome. The two gametes,
one from each parent, are now combined to form the genome of the
offspring. Each of its strands was formed out of a different set of
genes. The next stage of the reproduction process is the introduction
of $M_a$ independent mutations in the newly generated genetic strands
of the age-structured part of the genome. In this kind of model it is
normal to consider only the possibility of harmful mutations, because
of their overwhelming majority in nature, and this is what is done for the 
age-structured portion. For the non-structured portion, mutations are added 
independently of their harmful or beneficial effect, with some probability 
$M_p$ per string. The gender of the newborn is then randomly selected,
with equal probability for each sex. Figure \ref{schema} is a
schematic representation of the entire reproductive process.

The passage of time is represented by the reading of a new locus in the 
age-structured portion of the genome of each individual in the population, and the 
increase of its age by one. After having accounted for the selection pressure of a
limiting number of effective harmful mutations and the stochastic action of the 
fitness function, females that have reached the minimum age for reproduction 
generate a fixed number of offspring. The simulation runs for a pre-specified 
number of time steps, at the end of which averages are taken over the population. 
Typically, measures are taken for the age structure of the population, such as the 
number of individuals and probability of survival and death by genetic
causes for each age group, as well as for the genetic composition
distribution. We are particularly interested in the equilibrium
distribution of the phenotypes in the population, also averaged at the
end of each run.

\section{Simulation results}

We will relate our results qualitatively to the observations cited in the 
introduction made on the evolutionary studies of the Galapagos finches. Thus, we 
will have in mind a picture of our population as composed of ground finches, 
feeding primarily on seeds. The phenotype trait represented by our model describes 
the size and strength of their beaks, and the fitness measures their adequacy to 
the kinds and sizes of seeds available in the ecology. 

\begin{figure}
\centerline{\psfig{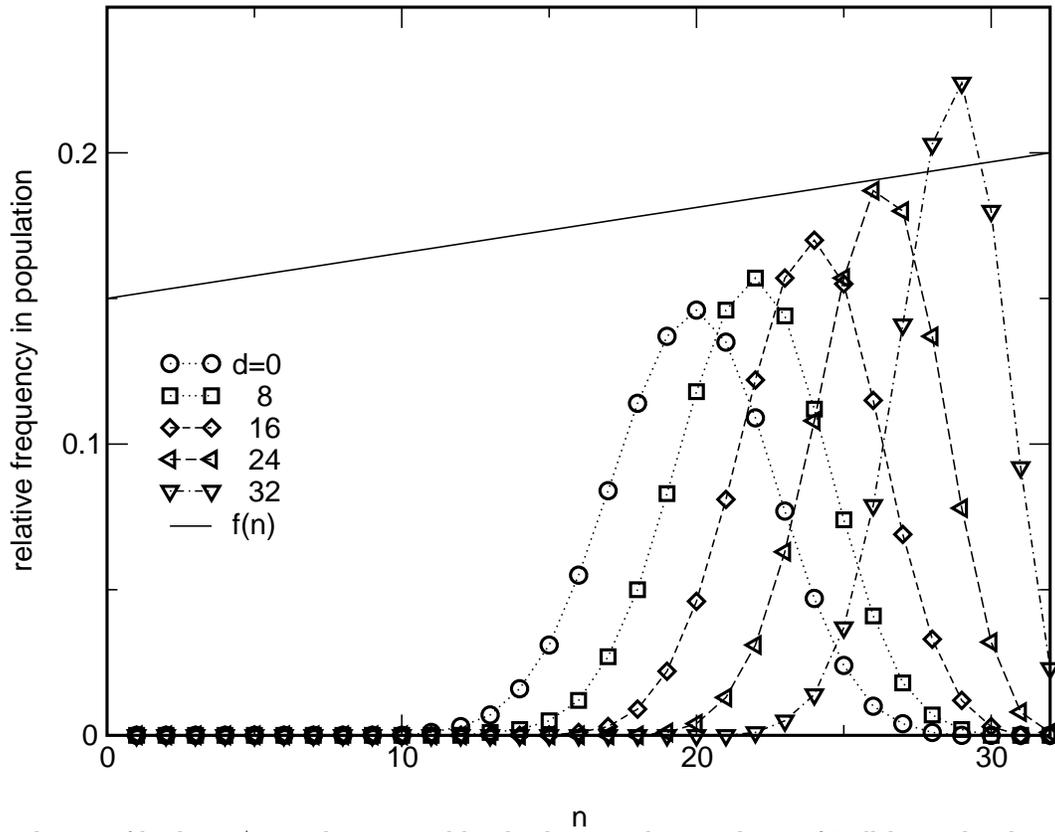}}
\caption{Distribution of beak size/strength, measured by the dimensionless number 
$n$ of $1$ alleles in the phenotype, over the population. Results are shown for 
various values of $d$, the number of loci where the $1$ allele is dominant. As $d$ 
increases, so does the mean value of $n$, whereas the width of the distribution 
decreases. When $d=32$ and the $1$ allele is dominant in all loci, the distribution 
approaches a narrow bell-shaped curve peaking at a value close to $32$. Also shown 
is the fitness function used in the runs, rescaled by a constant. All runs were 
$50 \ 000$ Monte Carlo time steps long, with a boost $A=128$, representing a weak 
selection. Other relevant parameters were:\\
$400 \ 000$ for $C$, the conventional carrying capacity of the standard Penna 
model;\\
$M_a = 1$; $M_p = 0.01$.}
\label{distdom}
\end{figure}

Let us start with results concerning the equilibrium configuration of the single 
trait that characterizes the fitness of an individual's phenotype. For this study, 
the fitness function is 
\begin{equation}
f(n) = 1 - \frac{32 - n}{A};
\label{fitness}
\end{equation}
$n$ is the number of bits set to $1$ in an individual's phenotype, and $A$ is a 
boost factor that can be used to control the intensity of the selective 
pressure; this intensity is a decreasing function of $A$. This fitness function 
expresses an environment where the ideal phenotype would be composed entirely 
of bits $1$. For instance, it can represent an environment in which the 
availability of edible seeds is a decreasing function of their size, while the 
number $n$ is a measure of the size of the beak of an individual. In this ecology, 
a well-fit individual would have a large beak, allowing him to take advantage of 
the distribution of seeds size. The simulation starts with an ill-fit population, 
in which all the individuals have the smallest beak possible; in our 
representation, their phenotypes are homozygotes $00$ in all loci. Alternatively, 
this can be expressed as a distribution of beak sizes delta-like peaked at $n=0$. 
Figure \ref{distdom} shows the equilibrium distribution reached after $50 \ 000$ Monte 
Carlo steps, for a weak selective pressure, corresponding to a broad distribution 
of seed sizes and, in our representation, to a large value of the boost $A=128$. We 
plot the distribution of beak sizes for various values for the number of loci $d$ 
in which the well-fit ($1$) allele is dominant in the phenotype. In all cases, the 
distribution shows a bell-shaped Gaussian-like aspect in which the peak position and 
amplitude increase with $d$, while the width shows a decreasing pattern. This result 
is easy to understand: because of the availability of large seeds, represented here 
by a fitness function that increases with $n$, the population is driven to a pattern 
of larger beaks. The combined effects of mutation pressure and heterozygose avoids 
its collapse into a delta-like distribution from which it evolved; the width of the 
distribution increases with the mutation rate. In a strongly selective ecology, 
represented by a smaller value of $A$, the pattern is essentially the same. The 
peaks are shifted to larger values of $n$ and the widths decrease.

\begin{figure}
\centerline{\psfig{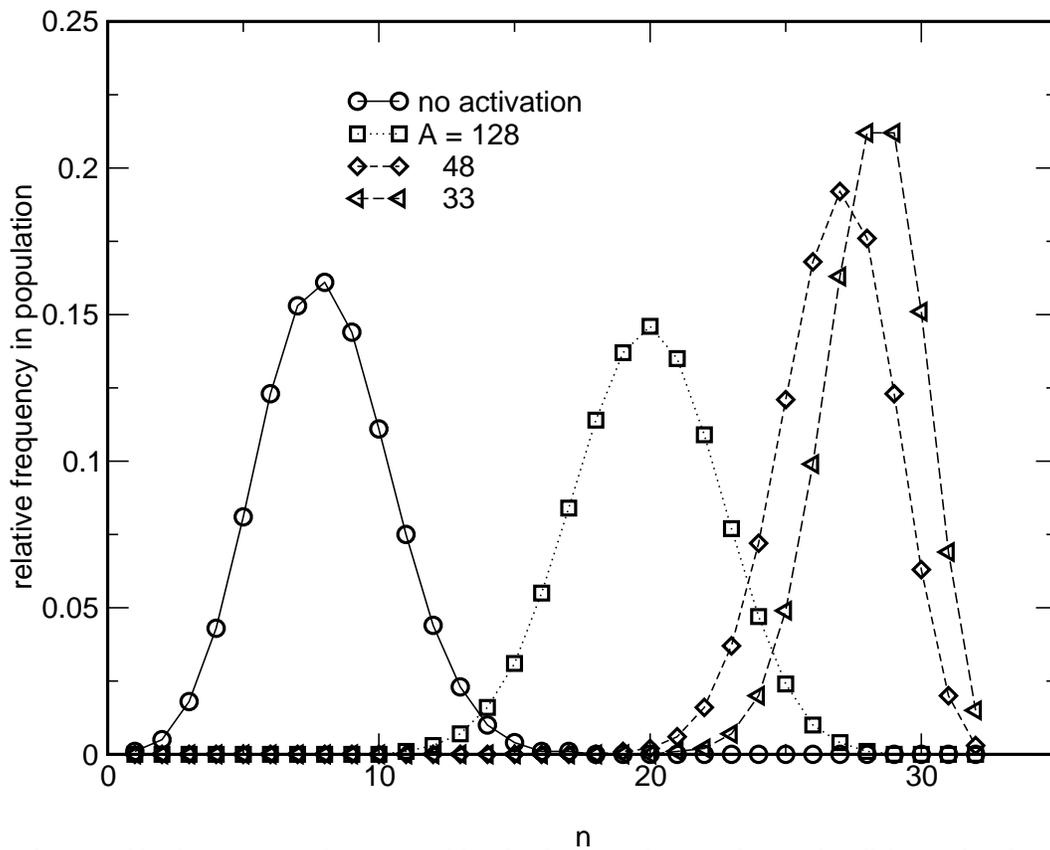}}
\caption{Distribution of beak size/strength, measured by the dimensionless number 
$n$ of $1$ alleles in the phenotype, over the population. Results are shown for 
various values of $A$, representing the intensity of the selection pressure, for a 
fixed value of $d=0$, meaning that the allele $1$ is recessive in all loci. The 
shift of the distribution to larger values of $n$ when the selection pressure gets 
stronger is clear. The parameters of the simulations were the same as in Figure 
\ref{distdom}}
\label{distboost}
\end{figure}

The shift in the distributions caused by stronger selection is shown in Figure 
\ref{distboost}. We compare results obtained for a fixed value of $d=0$, where the 
allele $1$ is recessive in all loci. When there is no activation of phenotype 
selection, which can be translated into a uniform distribution of seed sizes, the 
broad pattern of beak sizes is a result of random genetic drift. If $d=16$, the 
equilibrium distribution peaks at $n=16$. When phenotype selection is activated, 
the equilibrium distribution tries to follow the availability of seed sizes and 
its peak moves to larger values of $n$, while its width shrinks. It is worthwhile 
to remark that even in a very strong selective ecology ($A=33$), the distribution 
still retains some broadness, representing a diversity of morphologies also 
observed in real populations.

\begin{figure}
\centerline{\psfig{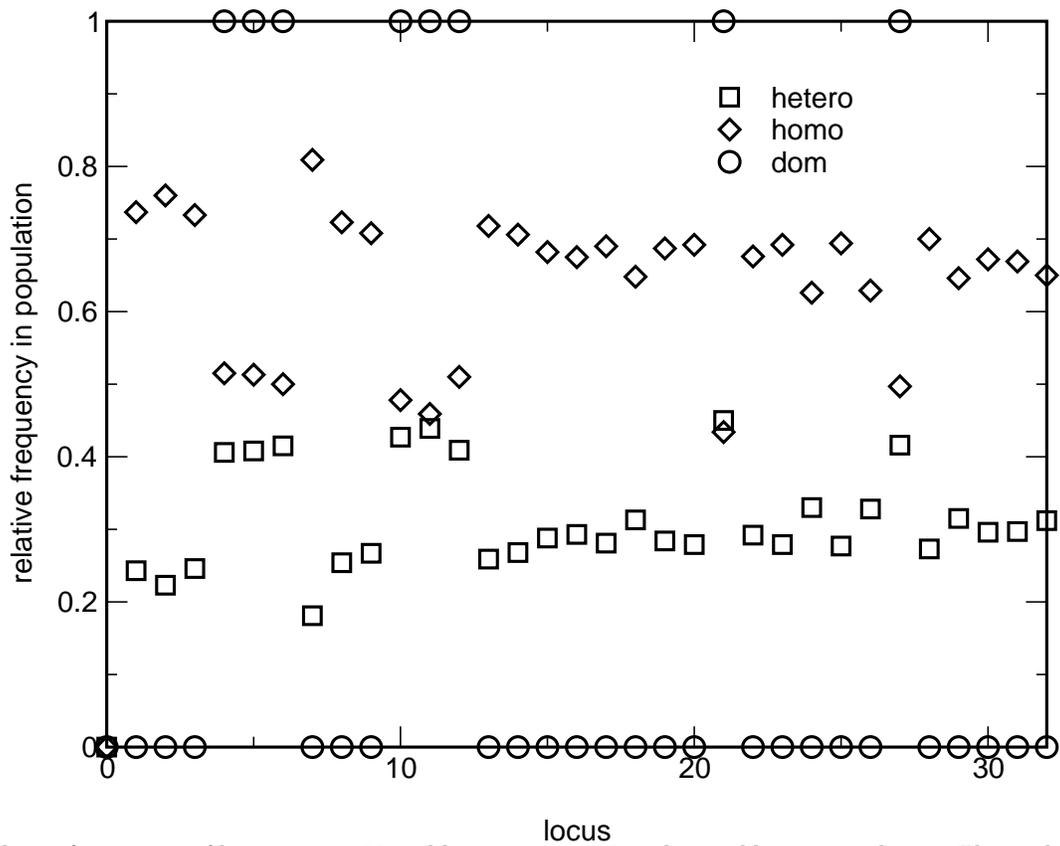}}
\caption{Relative frequencies of homozygotes $11$ and heterozygotes $10$ in the 
equilibrium population. The circles indicate, for each locus, which is the 
dominant allele. This run had the same parameters as the ones from which Fig 
\ref{distdom} was extracted, except for the boost $A=48$.}
\label{homo}
\end{figure}

We examine next the pattern of homo/heterozygose generated in the phenotype of the 
population. Figure \ref{homo} was obtained for a medium range selective pressure 
$A=48$. The circles indicate what is the dominant allele in each locus. One can 
immediately identify that the selection pressure pushes the loci in which the 
allele $0$ is dominant to homozygose $11$, while on sites where $1$ is dominant 
hetero and homozygose are in fact indistinguishable.

\begin{figure}
\centerline{\psfig{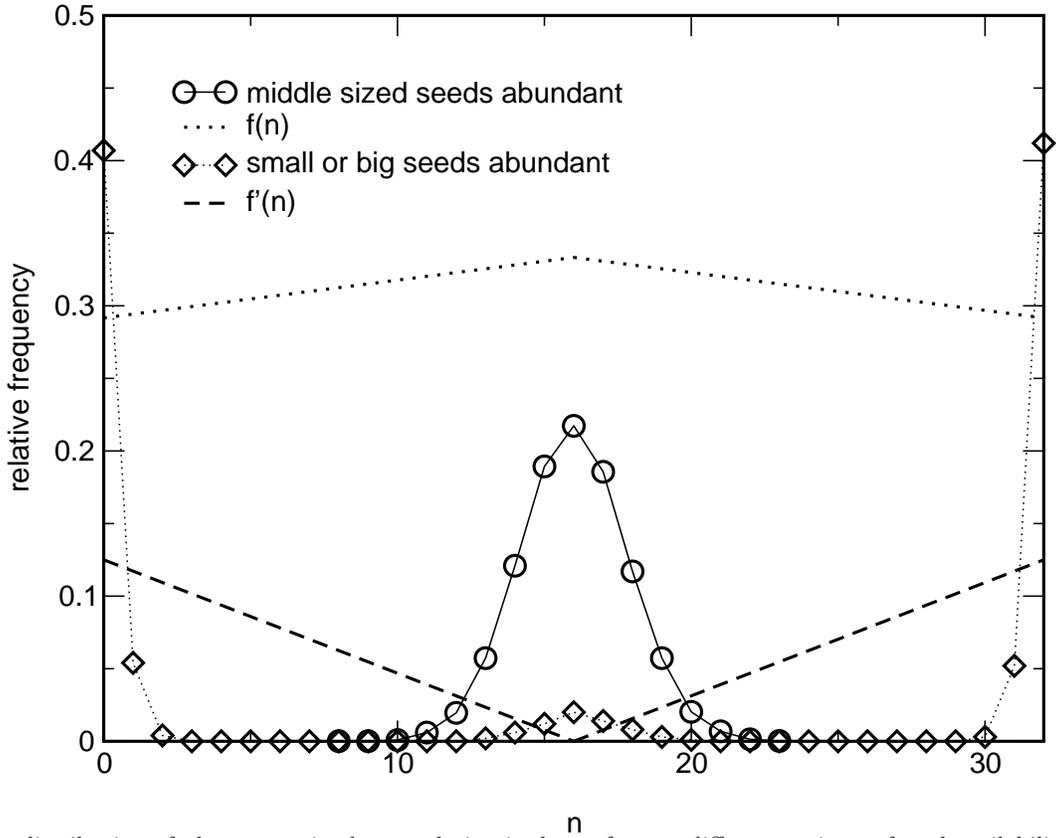}}
\caption{The distribution of phenotypes in the population is shown for two 
different regimes of seed availability. The circles correspond to the equilibrium 
population in a situation in which seeds are available for a broad distribution of 
sizes, peaked at $n=16$; the corresponding fitness function $f(n)$, adequately 
rescaled to fit in the graph, is shown for a comparison. The two-peaked phenotype 
distribution corresponding to the diamonds sets in after a climate-induced change 
in the food supply alters the fitness function to $f'(n)$. The population splits 
into a (reversible) polymorphism, with different beak sizes. Since there is no 
reproductive isolation, mating between birds feeding on different niches generate 
offspring with medium-sized beaks, represented by the small bump at $n=16$. The 
parameters of the simulation are the same as in the preceding figures, except 
where otherwise stated in the text.}
\label{poly}
\end{figure}

The establishment of a stable polymorphism is the theme of Figure \ref{poly}. In the 
simulation, the initial population was again a small sized beak one, now immersed 
in an ecology with a broad distribution of edible seed sizes available, peaking 
at middle-sized seeds. The fitness function that represents this situation is
\begin{eqnarray}
\nonumber
f(n) = 1 - \frac{16-n}{A}, \,\,\, n < 16\\
 = 1 - \frac{n-16}{A}, \,\,\, n \ge 16
\end{eqnarray}
The population evolves for $20 \ 000$ Monte Carlo steps, when a
snapshot of the phenotype distribution is taken. In agreement with the
result shown in Figure \ref{distdom}, this distribution is again
bell-shaped, with its peak now located at $n=16$, corresponding to 
middle-sized beaks. Because mutations can both increase or decrease
the beak size, as opposed to the case where the peak of the fitness
function was at the largest possible extreme, and because the number of 
loci where each allele is dominant is the same, corresponding to a
value $d=16$ for the dominance parameter, there is no inherent bias 
to the equilibrium distribution; its peak can sit at the same position
as the one for the fitness function, as Figure \ref{poly} shows. After
$20 \ 000$ time steps, there is a sudden change in the pattern of seed 
availability, perhaps because of a variation in the rainfall regime
caused by some global climate oscillation, whose effect is to decrease
the availability of middle-sized seeds. The fitness function that 
expresses this new pattern is, for instance,
\begin{eqnarray}
\nonumber
f'(n) = 1 - \frac{A-16+n}{A}, \,\,\, n < 16\\
 = 1 - \frac{A-n+16}{A}, \,\,\, n \ge 16
\end{eqnarray}
The evolution of the population resumes with this new fitness function, and with 
intra-specific competition within each niche. This is represented in the model by 
letting the mean-field $V(P)$ take into account {\it only the individuals that 
feed on a particular kind of seed}. Thus, finches with small (large) beaks, 
represented by having $n<(>)16$, and feeding on small (large) seeds, will compete 
only against those feeding on the same niche. To be specific, the simulation now 
computes at each time step $P_m = \Sigma_{population} \theta(16-n)$ and 
$P_M = \Sigma_{population} \theta(n-16)$, where $\theta(k)$ is the step function 
with $\theta(0)=1/2$. To determine the probability of death by intra-specific 
competition for an individual, its beak size, which is proportional to the number 
$n$ of $1$-bits set in its phenotype, is used to decide in which niche it is 
competing. 
If $n<(>)16$, its death probability will be $V(P_m)/f'(n)$ 
($V(P_M)/f'(n)$). These elements alone force evolution to give rise to a 
polymorphism, shown in Figure \ref{poly} as a resulting $2$-peaked equilibrium 
distribution of beak sizes. This polymorphism is reversible: if, in a subsequent 
time step, the pattern of availability of edible seeds reverts to its original 
configuration, so does also the distribution of beak sizes. This is in complete 
agreement with the field observations of stable polymorphism in the Galapagos 
ground finches \cite{grant,lack}.

If such a stable polymorphism is not challenged by new climate variations for some 
time, it can induce speciation and generate two reproductively isolated segments, 
even without geographical barriers. To simulate this effect, we introduced in our 
model a gene that determines mating selection pattern. The initial population has 
a panmictic behavior, expressing no preferences related to the phenotype of the 
partner. When a female is ready to mate, she chooses a partner according to the 
expression of this gene. A randomly selected male in the population, to be accepted 
as a partner, has to either feed on the same niche, in which case the mating 
selection gene becomes irrelevant, or, if he feeds on a different niche, both 
parents have to be non-selective in their mating preferences for reproduction to 
occur. The offspring inherit the mating preferences of either the mother or the 
father, randomly selected at birth, and this gene can also suffer a mutation in 
either direction with some probability. The resulting distribution of this gene 
among the population is shown in table \ref{mate}. After the double peaked 
polymorphism is established, and if the distribution of edible seeds does not 
change for some time, reproductive isolation sets in.

\section{Conclusions}

The addition of a multi-locus non-structured part to the
age-structured genome of the Penna model, subjected to the same 
representation of evolution dynamics, allows a substantial broadening
of its scope. The role of genes with age-independent effects can now
be studied through the patterns that they generate in evolutionary 
populations. A particularly interesting side benefit is to get rid of
the unmotivated random effect of the Verhulst dagger and substitute it by 
inter-specific competition with outcome based on the genetic load of the 
individuals. With the help of this broadened model we studied the 
distribution of a single phenotypical trait among the population when
it evolves on a single-peaked linear fitness landscape. The
simulations exhibited its Gaussian-like character with mean and width
that depend in a clear way on the various parameters of the modified
model. The distribution of homo and heterozygote loci of the
non-structured genome over the population also shows patterns that are 
easy to explain and to relate to the parameters. The present study
opens the path for a variety of new applications of age-structured
microscopic models as it establishes their basic properties.

As a hint of the scope of these applications, we undertook the
simulation of the effects of a varying pattern of food supply on the 
phenotype distribution . Our motivation were the reports from field 
observations of the variability of the beak morphology of the ground
finch in the Galapagos Islands as a response to changes in seed
availability due to seasonal oscillations in rainfall regime. The
results of the simulations show a striking parallelism with those 
observations, in that a polymorphism is generated as soon as the
fitness landscape switches to a double-peaked pattern. Since modern 
theories on speciation seem to indicate that polymorphism is a
necessary first step, we proceeded by introducing a gene that controls
mating preferences. Mutation and evolution under a continued double-peaked 
fitness landscape were enough to generate reproductive isolation and 
speciation as a result. Speciation without geographic isolation has
proven to be a difficult concept in evolution theory so far, but our
simulations support a scenario in which its onset arises as a
necessary evolutionary consequence of a continued double-peaked
fitness landscape. 

\section*{Acknowledgments}

We thank D. Stauffer for a critical reading of the manuscript. 
Research by J.S.S.M. was supported as a Visiting Fellow by CIRES, University 
of Colorado at Boulder. SMO thanks the Brazilian agencies CNPq, CAPES and 
FAPERJ for partial financial support.

\newpage

\newpage

\newpage

\newpage

\newpage

\begin{table}
\caption{Fraction of the population with reproductive isolation. Mutations of 
mating preferences at birth occur with a probability $p_{mate}=0.001$. The results 
are averages over $10$ runs.}
\label{mate}
\begin{tabular}{l|c|c}
& males & females \\
\tableline
small beaks/broad distribution of seeds & 0.003 & 0.003 \\
small beaks/two-peaked distribution of seeds & 0.939 & 0.940 \\
large beaks/broad distribution of seeds & 0.003 & 0.003 \\
large beaks/two-peaked distribution of seeds & 0.952 & 0.954 \\
\tableline
\end{tabular}
\end{table} 

\end{document}